# ALMA Memo No. 613
# ALMA FITS header keywords:
# a study from the archive User perspective.

Authors: Elisabetta Liuzzo, Marcella Massardi, Kazi L. J. Rygl, Felix Stoehr, Andrea Giannetti, Matteo Bonato, Sandra Burkutean, Anita Richards, Mark Lacy, Jan Brand

**Summary**

ALMA products are stored in the Science Archive in the form of FITS images. It is a common idea that the FITS image headers should collect in their keywords all the information that an archive User might want to search for in order to quickly select, compare, or discard datasets.
With this perspective in mind, we first present a short description of the current status of the ALMA FITS archive and images. We realized that at the moment most of the parameters that could be useful for a general User are still missing in the archived data. We then provide a CASA task generating the image header keywords that we suggest to be relevant for the scientific exploitation of the ALMA archival data. The proposed tool could be also applied to several types of interferometer data and part of it is implemented in a web interface. An example of the scientific application of the keywords is also discussed.

The organization of the Memo is as follows:
**Contents**

## 1. Scope

Images (maps and cubes) in the ALMA Science Archive (ASA) are stored in the Flexible Image Transport System (FITS) [1] format which is the standard archival data format for astronomical data sets providing a general way to encode both a definition of the data in the header and the data themselves in an operating system independent format.

In this document, we focus on the FITS headers. It is commonly accepted that a general image header should fully describe the image itself summarizing its content, the production process (i.e., data acquisition and calibration), and additional information to make it accessible to the common image analysis tools. For archived images, the FITS header should then be a repository of this information. For each ASA image, all the necessary details of the observation and reduction process that generated it (including the raw data, the calibration tables and the script that generated it) are accessible through the ASA itself. However, to date, to infer them it is typically necessary to download all the raw data, run the calibration scripts with the proper Common Astronomy Software Applications (CASA[2]) version and consult the calibrated Measurement Set (MS) data.
Moreover to extract other quantities, it is necessary to open the image and measure them. All these processes could be tedious and long, and the results could be inhomogeneous if they have to be done on a large number of images.

We foresee an ASA that collects images with informative FITS headers that (even if not due to be science-ready by themselves) are fully representative of the content and the processes applied to the data and provide a satisfactory indication of the quality of the achievable scientific results.
In this perspective *the FITS image header should collect all the information that an archive User might want to search for in order to quickly select a dataset, compare several datasets, or discard useless datasets.*

We start giving an update of the current status of the FITS headers of the ALMA archived images (Sect. 2). Hence, in addition to the established keywords with a well defined meaning identified by the FITS Standard 3.0 documents (Pence et al. 2010), and the updated 4.0 version [3], we suggest a list of new keywords (Sect. 3) that, on the basis of our experience, could be useful to the general ASA User for purposes like those described above.
We thus present in Sect. 4 the ALMA Keywords Filler CASA (AKF) task that generates the image header keywords presented in the current document. Their values could be either obtained from the image itself or from the calibrated MS according to the given definitions. We introduce in Sect. 5 the AKF web interface for the image-description keywords, KAFE.
In order to drive our suggestions for a comprehensive FITS header we have investigated some scientific applications (Sect. 6) that could benefit from an easily accessible, fully documented keywords set. We also provide the updated list of recommended keywords (Appendix A) for the ASA FITS image header (including those identified by the FITS standard 4.0 and those suggested in the present document), and an example of the resulting FITS header format (Appendix B). Finally in the Appendix C, we provide an example of the use of the AKF CASA task to produce scientific plots.

---

1 https://fits.gsfc.nasa.gov/fits_standard.html
2 https://casa.nrao.edu
3 https://fits.gsfc.nasa.gov/standard40/fits_standard40aa.pdf

## 2. Current status of ALMA Science archive

The structure of the data trees stored in the archive reflects the project processing structure[4]. An ALMA project is split into Science Goals (SG, the minimum observation settings and targets to reach a scientific purpose), each of which is translated at the observing stage into a Group Units and split into Member Observing Units (MOUs) separating the different settings of the array, each of which is translated into code instruction to the array called Scheduling Blocks (SBs). In order to maximize the efficiency of the telescope dynamical schedule and, as a consequence of the observations, SB are limited in time and repeated as many times as needed to reach the Principal Investigator (PI) requested sensitivity and resolution: each SB repetition is called Execution Block (EBs) and it constitute an independent observing run enclosing its own calibrating source observations. Hence, each analyst should calibrate each EB of a given MOU and combine them all to produce the product images to estimate the reached sensitivity and angular resolution for a given observational setting in a SG: the quality assessment definition works at this level.

In ALMA, in fact, a layered quality assessment (QA) process[5] is applied to all the datasets: after checking the optimal telescope conditions for the data to be taken and stored on the telescope site (QA0 and QA1), the data are fully calibrated and a minimal imaging is performed to verify that the resolution and sensitivity requested by the PI are reached (QA2): in case of negative response additional EB are observed (if possible), otherwise, they are delivered to the PI either as "QA2-pass" or "QA2-semipass". Please notice that for manual assessors imaging of at least for one source in continuum and in one spectral line is requested, so that any additional produced image is done on the assessor goodwill/time/capability. Product images are not intended to be science ready (as calibrated data are), and it is expected that PIs or archive miners use them only as indicators of data quality. In the archive, raw data for each of the >6000 EBs observed so far are stored and properly linked to the project tree they belong to. Scripts to calibrate each EB resultant of the (manual- or, more recently pipeline-based) QA, as well as preliminary scripts for imaging of the whole MOU are stored as well, together with the images produced during the QA and a set of calibration diagnostic plots. Cycle 0 data constitute an exception, as only the raw data are stored and they are not organized according to the above described data tree.

Currently, an archive miner can choose for each MOU whether to download only the products (scripts, images and diagnostic plots), with a typical size of a few 100MB and/or the raw data that could reach size of several 100GB depending on the observing settings (number of antennas, frequency channels, EBs,...). The data download might thus take several hours to days. In order to complement or re-produce the product FITS images, once all the data are downloaded, calibration scripts should be run with the same CASA version that the analyst used to generate them (also this process might require hours to days). Only then, the User should use (and frequently improve or adjust) the distributed imaging scripts to produce the FITS images they need, sometimes to discover that the target they were looking for is undetected or for any reason, not observed.

---

4   **S**ee Sect. 8 in https://almascience.nrao.edu/documents-and-tools/cycle5/alma-technical-handbook/view
5   See Sect. 11 in https://almascience.nrao.edu/documents-and-tools/cycle5/alma-technical-handbook/view

## 2.1 Current status of ALMA FITS image header

In the following, we show one example of the FITS header of an archival Cycle 1 ALMA FITS image.

```
ALMA FITS header
SIMPLE  =                    T /Standard FITS
BITPIX  =                  -32 /Floating point (32 bit)
NAXIS   =                    4
NAXIS1  =                   84
NAXIS2  =                   80
NAXIS3  =                 3772
NAXIS4  =                    1
EXTEND  =                    T
BSCALE  =    1.000000000000E+00 /PHYSICAL = PIXEL*BSCALE + BZERO
BZERO   =    0.000000000000E+00
BMAJ    =    1.206100316999E-03
BMIN    =    7.315292685941E-04
BPA     =   -7.971664204539E+01
BTYPE   = 'Intensity'
OBJECT  = 'pi_Gru   '
BUNIT   = 'Jy/beam  '           /Brightness (pixel) unit
RADESYS = 'ICRS    '
LONPOLE =    1.800000000000E+02
LATPOLE =   -4.594799027778E+01
PC01_01 =    1.000000000000E+00
PC02_01 =    0.000000000000E+00
PC03_01 =    0.000000000000E+00
PC04_01 =    0.000000000000E+00
PC01_02 =    0.000000000000E+00
PC02_02 =    1.000000000000E+00
PC03_02 =    0.000000000000E+00
PC04_02 =    0.000000000000E+00
PC01_03 =    0.000000000000E+00
PC02_03 =    0.000000000000E+00
PC03_03 =    1.000000000000E+00
PC04_03 =    0.000000000000E+00
PC01_04 =    0.000000000000E+00
PC02_04 =    0.000000000000E+00
PC03_04 =    0.000000000000E+00
PC04_04 =    1.000000000000E+00
CTYPE1  = 'RA---SIN'
CRVAL1  =    3.356843454167E+02
CDELT1  =   -2.166666666667E-04
CRPIX1  =    4.300000000000E+01
CUNIT1  = 'deg     '
CTYPE2  = 'DEC--SIN'
CRVAL2  =   -4.594799027778E+01
CDELT2  =    2.166666666667E-04
CRPIX2  =    4.100000000000E+01
CUNIT2  = 'deg     '
CTYPE3  = 'FREQ    '
CRVAL3  =    3.423942288270E+11
CDELT3  =    4.883342805786E+05
CRPIX3  =    1.000000000000E+00
CUNIT3  = 'Hz      '
CTYPE4  = 'STOKES  '
CRVAL4  =    1.000000000000E+00
CDELT4  =    1.000000000000E+00
CRPIX4  =    1.000000000000E+00
CUNIT4  = '        '
PV2_1   =    0.000000000000E+00
PV2_2   =    0.000000000000E+00
RESTFRQ =    3.433000000000E+11 /Rest Frequency (Hz)
SPECSYS = 'LSRK    '            /Spectral reference frame
ALTRVAL =    7.909798029106E+05 /Alternate frequency reference value
ALTRPIX =    1.000000000000E+00 /Alternate frequency reference pixel
VELREF  =                  257 /1 LSR, 2 HEL, 3 OBS, +256 Radio
COMMENT casacore non-standard usage: 4 LSD, 5 GEO, 6 SOU, 7 GAL
```

```
TELESCOP= 'ALMA    '
OBSERVER= 'srams   '
DATE-OBS= '2013-10-08T02:46:00.960000'
TIMESYS = 'UTC     '
OBSRA   =     3.356843454167E+02
OBSDEC  =    -4.594799027778E+01
OBSGEO-X=     2.225142180269E+06
OBSGEO-Y=    -5.440307370349E+06
OBSGEO-Z=    -2.481029851874E+06
INSTRUME= 'ALMA    '
OBJECT  = 'pi_Gru  '
TELESCOP= 'ALMA    '
DISTANCE=     0.000000000000E+00
FIELD   = 'pi_Gru  '
ITER    =                    1
SPW     = '18      '
TYPE    = 'pbcorimage'
DATE    = '2016-09-16T12:02:40.934000' /Date FITS file was written
ORIGIN  = 'CASA 4.7.38256-REL (r38256)'
HISTORY CASA START LOGTABLE
HISTORY 2016-09-16T10:57:12 INFO SRCCODE='::setmiscinfo'
HISTORY Ran ia.setmiscinfo
HISTORY 2016-09-16T10:57:12 INFO SRCCODE='::setmiscinfo'
HISTORY ia.setmiscinfo(info={...})
HISTORY CASA END LOGTABLE
END
```

Investigating different ASA FITS images, we noticed that their FITS headers currently stored in the science archive contain keywords that can change on the basis of the CASA version or the procedures that produced them (manual or through pipeline).

Moreover, a lot of keywords lack a clear definition in the FITS header documentation (see orange lines). Other keywords appear to be redundant (see yellow lines).

Some additional keywords might be specifically requested by external analysis tools or to deal with data in other spectral bands.

In this document, we focus mostly on those that, according to our experience, could be of use for general radio to sub-mm interferometric scientific purposes.

## 3. FITS keywords for ALMA archive purposes

In the light of the current status of the ALMA archive, we describe in the this section the keywords that we suggest could be useful for the general ASA User, in addition to those identified by the FITS Standard v. 4.0 document.
We classified two main categories of such keywords:
- *data acquisition and reduction keywords:* these refer to the properties of the telescope during the observation and to the calibration process. Their values have to be extracted from the calibrated MS typically browsing the data tables, from the calibration products or from the proposal documents.
- *image description keywords:* these refer to the properties of the FITS image that hosts them.

For all the proposed keywords, we provide appropriate definitions: their values are extracted and calculated according to them (Sect. 3.1 and 3.2). In the Appendix A, we propose additional FITS keywords that we think will benefit a general ASA miner and that we plan to implement in our code (Sect. 4) in the next future. In the Appendix B we give examples of the format we recommend the FITS keywords should have.
The suffix *'K'* is added to distinguish the new keywords that we propose from those already present in the FITS headers.

### 3.1 Data acquisition and reduction keywords

KRATARG and KDECTARG
  Description: RA and DEC of the source as listed in the MS.
  Units: degrees
  Type: float
  Notes: in the presence of multiple MS with different source coordinates or mosaic, a list of RA and DEC is given

KUVRANGE
  Description: median, first, and third quartile of the UV length distribution
  Units: float
  Type: kilowavelengths
  Notes: in the presence of multiple MS, the total UV length distribution is the sum of each MS UV length distribution

KBAND
  Description: the receiver band used for the image
  Values: 'BX', where X= integer 3-10
  Type: character

KMINPRBL and KMAXPRBL
  Description: minimum and maximum projected baseline
  Units: m
  Type: float
  Notes: in the case of multiple MS, it calculates the minimum and maximum projected baseline of each MS and takes the minimum and maximum value among them. In the case of multiple

arrays, the minimum and maximum among the array is reported.

**KMAXANGS**

Description: the maximum angular scale resolved by the 12m, and 7m array
Units: arcseconds
Type: float
Notes: it is calculated as $37100/((kBNDCTR*10^{-9})* kMINPRBL)$. In the case of multiple arrays, the kMAXANGS of each array is listed. For kMINPRBL definition see above and for the kBNDCTR see below.

**KPADLIST**

Description: list of ALMA pads contributing to the data
Type: character
Notes: antennas with all data flagged are discarded. In the case of multiple MS, the final list contains all the pads, even if some are present only in one of the MS. In the case of multiple arrays, all the pads of each array is listed.

**KNANT**

Description: number of ALMA 12m, 7m, and ACA total power antennas contributing to the data.
Type: integer
Notes: In the case of multiple MS, the maximum number of antennas among each single MS is taken. Antennas with all data flagged are not considered.

**KDATEOBS, KDATEEND, KDATEAVG**

Description: start, end and mid-point of the observation
Units: year-month-dayThh:mm:ss (FITS ISO standards)
Type: character
Notes: they are defined from the first (min_obst) and last integration time (max_obst) and from the integration time 't'. For the starting time a 0.5*integration time is subtracted (KDATEOBS = min_obst – 0.5*int), while for the ending time 0.5*integration time is added (KDATEEND = max_obst + 0.5* int ) to reflect to precise beginning and end of the observation. Finally, the mid-point of the observations is the medium time from the start and the end (KDATEAVG= min_obst + max_obst /2).

**kMJDOBS, kMJDEND, kMJDAVG**

Description: Modified Julian Date (JD – 2,400,000.5) of start, end, and mid-point of the observation
Units: days
Type: float with a format F5.5
Notes: KMJDOBS, KMJDEND, KMJDAVG are the same as KDATEOBS, KDATEEND, KDATEAVG but in MJD format.

**3.2 Image description keywords**

KBNDRES
    Description: frequency resolution of the image defined as increment of frequency axis
    Units: Hz
    Type: float
    Notes: overlap with CDELTn (Ctype = FREQ)

KCHNRMS
    Description: the inter-quartile range of the pixel values in the image
    Units: Jy/beam
    Type: float
    Notes: for spectral line data, the inter-quartile range is taken considering the pixel values of all the channels together. In the case of polarization data, one value for each Stokes is provided.

KSPATRES
    Description: geometric average of the min and the max beam axes.
    Units: arcsecond
    Type: float
    Notes: for spectral line and polarization data, it loops over the channel and Stokes axes, takes the mean of min and max beam axes among all channel and Stokes values and then it calculates the geometric average of min and max beam axes.

KSTOKES
    Description: list of image Stokes parameters
    Value: either "I" or "I, Q, U, V"
    Type: character
    Notes: it covers the presence of total intensity I images or the full stokes parameters I, Q, U, V, while the cases of "LL, RR, RL, LR," and "XX,YY, XY, YX" are not yet implemented.

KDATAMIN and KDATAMAX
    Description: minimum and maximum values in the image
    Units: Jy/beam
    Type: float
    Notes: in the case of spectral line data, the minimum and the maximum among all the channel is taken. In the case of polarization data, one value is given for each Stokes.

KDYNRNG
    Description: estimation of image dynamic range for each Stokes defined as
    KDATAMAX /KCHNRMS
    Type: float
    Notes: for polarization data, one value is provided for each Stokes. As one estimation of KCHNRMS (see its definition above) is given for all channel, one value of KDYNRNGE for each Stokes is given.

KBNDCTR
    Description: the center frequency of the image calculated as ½ (the reference value for the frequency axis * number of channels) .
    Units: Hz
    Type: float

KBNDWID
    Description: the effective bandwidth of the image calculated as the increment in the frequency axis * number of channels
    Units: Hz
    Type: float

KFLUXTOT
    Description: integrated flux of the source obtained masking the image below 3 * KCHNRMS
    Units: Jy
    Type: float
    Notes: for spectral line data, the sum of pixels with flux above 3*KCHNRMS of each channel is taken. In the case of polarization data, one value is given for each Stokes.

## 4. AKF: the ALMA Keywords Filler CASA task

A CASA task, the AKF, is build to generate and eventually ingest in the headers the FITS keywords discussed in the previous Sect. 3. The AKF is a Python-based script that exploits existing CASA tasks and toolkit[6]. It is downloadable (AKF.tar) from the Italian ALMA Regional Center web page http://www.alma.inaf.it/index.php/ALMA_FITS_Keywords

In the following, the instructions to install the task are described:
- check that astropy, numpy and pyfits are installed
- untar the AKF.tar file into a directory of your choice
  (e.g., YourHomeDir/ITALIAN_TOOLS)
- enter in that directory
  (i.e. `cd YourHomeDir/ITALIAN_TOOLS`)
- launch CASA and run "buildmytasks" in that directory
  (i.e. `os.system('buildmytasks')`)
- rename mytasks.py in ITtask_AKF.py
  (i.e. `mv mytasks.py ITtask_AKF.py`)
- run the ITtask_AKF.py file
  (i.e. `execfile('YourHomeDir/ITALIAN_TOOLS/ITtask_AKF.py')`)
- you should be able to run the new task in CASA just doing:
  `inp AKF` or `tget AKF`

The inputs of the task are shown in Fig. 1:
- `imName`: it is the input image to process and it is a mandatory parameter to provide;
- `kwdlist`: it consists in the list of keywords to be calculated. The allowed keywords are those of Sect. 3. If the list is not given (kwdlist=[]), all the implemented keywords will be calculated.
- `outfile`: the name of disk file to write the output could be given. If no outfile name is given, the results will be written only to the terminal.
- `include`: it allows to ingest the calculated keywords in the FITS headers. The default is false which means that no keywords will be added to the headers.

```
---------> inp(AKF)
#   AKF :: FITS keywords
imName           =          ''          # Input image name
kwdlist          =          []          # Select kwds: [] ==> all kwds will be
                                        #   calculated
outfile          =          ''          # Name of disk file to write output, ''==>to
                                        #   terminal
include          =          False       # Include the calculated kwds in the FITS
                                        #   header, True or False. Default is False
```

*Fig. 1: The AKF input parameters.*

It is worth to note that to process the *data acquisition and reduction keywords* (Sect. 3.1), it is required that (only) the MS(s) from which the input FITS image was produced is (are) in the same folder as the image. Since the program browses the MS tables, it can take some minutes to run (and up to ~30 min for the biggest datasets as large as 100 Gb). If the MS(s) file is (are) not present, and these keywords are requested, a terminal message is written (e.g. see Fig. 2).
The calculated keywords could be written only in the terminal (Fig.2) as well in the casalog (Fig. 3) or in a output file (Fig. 4) .

---
6  https://casa.nrao.edu/docs/CasaRef/CasaRef.html

```
Executing:  AKF()

Given kwds : []
Given image : /Users/eliuzzo/Job/ALMA/AKF/AKF/AKF_codes/test2/Mrk590.fits
Allowed kwds : ['KBNDRES', 'KBNDWID', 'KBNDCTR', 'KDATAMAX', 'KDATAMIN', 'KCHNRMS', 'KS
PATRES', 'KSTOKES', 'KDYNRNG', 'KFLUXTOT', 'KDECTARG', 'KRATARG', 'KUVRANGE', 'KDATEEND
', 'KDATEOBS', 'KDATEAVG', 'KMJDAVG', 'KMJDOBS', 'KMJDEND', 'KNANT', 'KPADLIST', 'KBAND
', 'KMAXANGS', 'KMINPRBL', 'KMAXPRBL']
No disk file will be written. Results will be shown on the terminal.
No keywords will be added to the FITS header.
All allowed kwds will be calculated
Warning: no field_name and/or spw are available. Some keywords could not be calculated
Warning: KDECTARG will not be calculated as no MS is given
Warning: KRATARG will not be calculated as no MS is given
Warning: KUVRANGE will not be calculated as no MS is given
Warning: KDATEEND will not be calculated as no MS is given
Warning: KDATEOBS will not be calculated as no MS is given
Warning: KDATEAVG will not be calculated as no MS is given
Warning: KMJDAVG will not be calculated as no MS is given
Warning: KMJDOBS will not be calculated as no MS is given
Warning: KMJDEND will not be calculated as no MS is given
Warning: KNANT will not be calculated as no MS is given
Warning: KPADLIST will not be calculated as no MS is given
Warning: KBAND will not be calculated as no MS is given
Warning: KMAXANGS will not be calculated as no MS is given
Warning: KMINPRBL will not be calculated as no MS is given
Warning: KMAXPRBL will not be calculated as no MS is given
KBNDCTR : 480995584264.0

KBNDRES : 16000072169.3

KBNDWID : 16000072169.3

KCHNRMS : {'I': 0.000663904967950657 01}

KDATAMAX : {'I': 0.006486979778856039}

KDATAMIN : {'I': -0.0049560186453163624}

KDYNRNG : {'I': 9.7709462829899572}

KFLUXTOT : {'I': 0.27562828864499983}

KSPATRES : 0.11

KSTOKES : I
```

*Fig. 2: Example of the messages written in the CASA terminal when all the implemented keywords are required for calculation but no MS file is provided.*

We caveat the reader that not only the FITS Standard v. 4.0 keywords but also others could be present in the headers of ALMA archival images as result of e.g. ALMA pipeline data reduction and we do not at any stage recommend to remove or overwrite them. This is why the AKF code is build to add

keywords of Sect.3 but not to remove or overwrite the already present FITS keywords in the header.

```
INFO       imstat::::   Warning: KDECTARG will not be calculated as no MS is given
INFO       imstat::::   WARNING: KRATARG will not be calculated as no MS is given
INFO       imstat::::   Warning: KUVRANGE will not be calculated as no MS is given
INFO       imstat::::   Warning: KDATEEND will not be calculated as no MS is given
INFO       imstat::::   Warning: KDATEOBS will not be calculated as no MS is given
INFO       imstat::::   Warning: KDATEAVG will not be calculated as no MS is given
INFO       imstat::::   Warning: KMJDAVG will not be calculated as no MS is given
INFO       imstat::::   Warning: KMJDOBS will not be calculated as no MS is given
INFO       imstat::::   Warning: KMJDEND will not be calculated as no MS is given
INFO       imstat::::   Warning: KNANT will not be calculated as no MS is given
INFO       imstat::::   Warning: KPADLIST will not be calculated as no MS is given
INFO       imstat::::   Warning: KBAND will not be calculated as no MS is given
INFO       imstat::::   Warning: KMAXANGS will not be calculated as no MS is given
INFO       imstat::::   Warning: KMINPRBL will not be calculated as no MS is given
INFO       imstat::::   Warning: KMAXPRBL will not be calculated as no MS is given
INFO       AKF::::      Input parameters:
INFO       AKF::::      imName = /Users/eliuzzo/Job/ALMA/AKF/AKF/AKF_codes/test2/Mrk590.fits
INFO       AKF::::      kwdlist = ['KBNDRES', 'KBNDWID', 'KBNDCTR', 'KDATAMAX', 'KDATAMIN', 'KCHNRMS',
INFO       AKF::::      No disk file will be written. Results will be shown on the terminal.
INFO       AKF::::      No keywords will be added to the FITS header.
INFO       AKF::::      KBNDCTR : 480995584264.0
INFO       AKF::::      KBNDRES : 16000072169.3
INFO       AKF::::      KBNDWID : 16000072169.3
INFO       AKF::::      KCHNRMS : {'I': 0.00066390496795065701}
INFO       AKF::::      KDATAMAX : {'I': 0.006486979778856039}
INFO       AKF::::      KDATAMIN : {'I': -0.0049560186453163624}
INFO       AKF::::      KDYNRNG : {'I': 9.7709462829899572}
INFO       AKF::::      KFLUXTOT : {'I': 0.27562828864499983}
INFO       AKF::::      KSPATRES : 0.11
INFO       AKF::::      KSTOKES : I
INFO       AKF::::      ##### End Task: AKF                          #####
INFO       AKF::::+     ##########################################
```

*Fig. 3: Example of the AKF CASA log when all the implemented keywords are asked to be calculated but no MS file is provided.*

```
KBNDCTR 480995584264.0
KBNDRES 16000072169.3
KBNDWID 16000072169.3
KCHNRMS {'I': 0.00066390496795065701}
KDATAMAX        {'I': 0.006486979778856039}
KDATAMIN        {'I': -0.0049560186453163624}
KDYNRNG {'I': 9.7709462829899572}
KFLUXTOT        {'I': 0.27562828864499983}
KSPATRES      0.11
KSTOKES I
```

*Fig. 4: Example of the output file provided by the AKF task when outfile is given in the case of kwdlist =[], and no MS file is provided in the same FITS folder.*

**5. KAFE: the AKF web interface for image-description keywords**

The part of the AKF code related to the image description keywords (Sect. 3.2) is implemented in the Key-analysis Automated FITS-images Explorer (KAFE - Burkutean et al. 2018), a Python tool suite that exploits CASA tasks and toolkit.
Even if originally conceived in the framework of our ALMA activities, KAFE is well suited to analyze FITS images produced by most of the major radio to submm interferometric facilities.
In particular, KAFE offers a web interface (see Fig. 5) that allows:

- to fill the header of the input FITS images with products of the image post-processing (e.g. total flux, dynamic range, rms). This offers quick-look catalogues, and a fast, homogeneous and coherent comparison tool for image sets (see Sect. 6 for scientific applications);
- to provide advanced diagnostic plots (e.g. moments maps, spectrum, polarization vectorial images, SEDs, light curves …);
- to cross-match the image with the most widely used astronomical catalogues and databases;
- to exploit some visualization tools (e.g. 3-color images, Mollweide representation).

Allowing a quick comparison of the data contents in the form of advanced image metadata as well as diagnostic plots, KAFE is then suitable for all the major data mining purposes.
We remind that first release of the KAFE implements only the AKF code to derive keywords related to the images (Sect. 3.2) . In later versions, also the keywords related to the MS properties (Sect. 3.1) will be provided.

*Fig. 5: A screenshot of the KAFE web interface.*

**6. Science cases for use of keywords in archived FITS images**

As astronomy continues to move towards multi-wavelength and data-driven science, issues of data provenance become of vital concern. Many future User cases for ALMA data will involve the download of FITS images from the archive through protocols such as the Virtual Observatory, where the User will receive the data file with essentially no other piece of information. To make the best possible use of such data in a publication, the metadata in the file must contain sufficient provenance information to permit the replication of the published results, sufficient characterization to allow meaningful statements about the nature of the observations, and sufficient attribution to guarantee that organizations and individuals are given due credit.

FITS is the only data standard used commonly in all fields of astronomy. Thus, the most convenient way to supply these data is via keyword-value pairs in the FITS header. The use of this technology ensures that such data can be easily machine readable, for example into databases and/or Python dictionaries. In addition to that, keyword values could be used for image selection, comparison and statistical analysis with direct scientific exploitation.
Here we summarize some of the most relevant examples of archived FITS image keywords exploitation, and how they could benefit of the AKF tool.

The AKF code is particularly useful to compare image products, for example to evaluate image quality or identify the images to be selected for User scientific purposes. It is important to stress that to exploit AKF for publications we recommend to apply the codes to calibrated data (tagged as science-ready) or re-imaged according to the User's needs.

Some examples of scientific application for the image-description part of the code are presented in Burkutean et al. 2018 and Massardi et al. 2019. Here we present a science case that exploits also the data acquisition keywords: the case of light curves and source variability analysis.

Light curves show the brightness of an object over a period of time and they are a simple, but useful tool to understand processes at work within the objects, such as novae, supernovae, variable stars but also blazars, or extragalactic objects in general. To construct them, it is necessary to retrieve an adequate number of images, typically through archive mining processes, to derive the brightness or flux over time. We will show how the AKF code could help in automatically producing light curves in a short time for a huge amount of observations.

Here we discuss in particular the case of blazar. For a proper characterization of the properties of this source population, the analysis of light curves on statistical significant samples is  mandatory.
Blazars, as a class of Active Galactic Nuclei (AGN, Urry & Padovanni 1995), are characterized by strong variability in all wave-bands. Additionally, a broad range of variability time scales is observed, ranging from minutes, as in the cases of PKS 2155-304 (Aharonian et al. 2007) and PKS 1222+216 (Tavecchio et al. 2011), to months (e.g. Abdo et al. 2010). In particular, the very short variability time scales are puzzling, since their emission should be generated in emitting regions much smaller than the event horizon of the AGN black hole, which, instead, should be the lower limit on the jet width size.
Blazars emitting at high energy are peculiar sources for which the emission mechanism and site of the gamma-ray signal are not fully understood. In this case, the cross-correlation of the radio/millimeter emission with the gamma-ray one is crucial to test the particle acceleration models and the origin of the high energy emission.
The high energy emission and the erratic, rapid and large-amplitude variability observed in all accessible spectral regimes (radio-to-gamma-ray) are two of the main defining properties of blazars (e.g. Ulrich et al. 1997; Webb 2006). The entire non-thermal continuum is believed to originate mainly in a relativistic jet, pointing close to our line of sight. Studies of variability in different spectral bands and correlations of multi-waveband variability patterns allow us to shed light on the physical processes in action in blazars, such as particle acceleration and emission mechanisms, relativistic beaming, origin of flares and size, structure and location of the emitting regions that could be complemented with the high resolution- sensitivity ALMA maps.
In Fig. 6 we report an example of light curves derived using the AKF task for the blazar source J0635-7516 (Bonato et al. 2018). In the Appendix C, we present the Python code that exploits the AKF task, through the calculation of the KFLUXTOT and KMJDAVG keywords (and kìKCHNRMS for the error estimates), to produce those light curves.

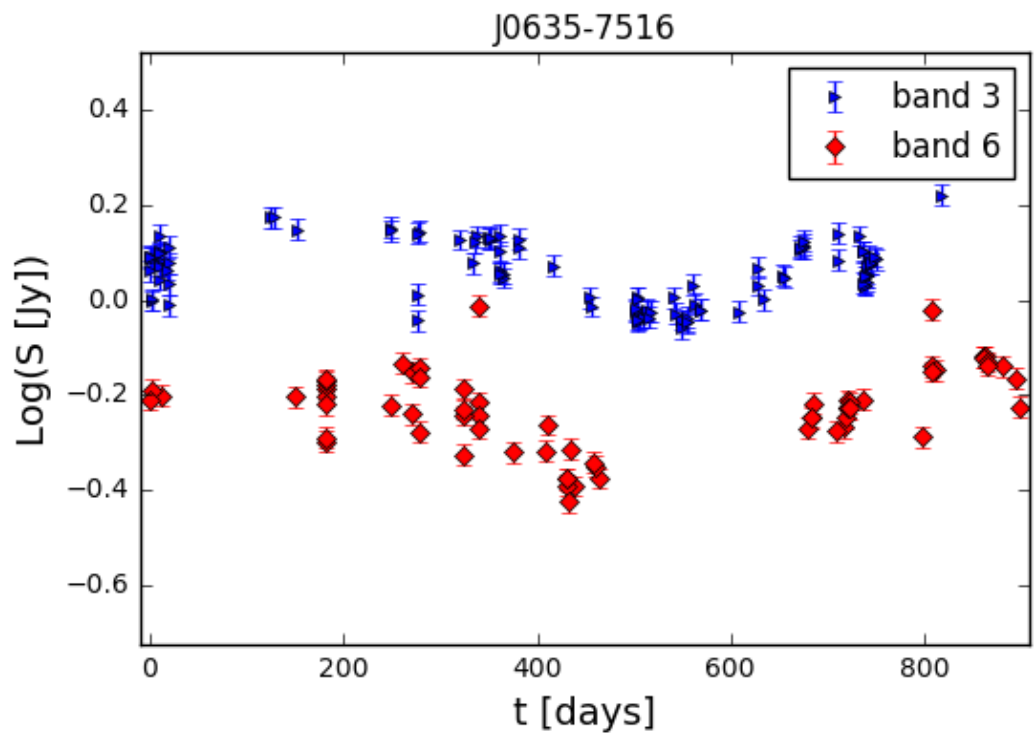

*Fig. 6: Example of blazar light curves obtained using the AKF task: the case of J0635-7516 (Bonato et al. 2018.)*

**Appendix**

**A - Recommended list of keywords**

In the following, we first list the existing FITS Standard 4.0 keywords and then those we suggest will be useful to include in the metadata of images as suitable for a general archive User.
In blue are the keywords already provided by the AKF code or by the FITS Standard 4.0 or already present in ALMA FITS headers with clear definition.
In red we mark those that are under construction by the AKF project (and that will be distributed in future versions of the code) and those for which the information is currently not derivable directly by the FITS or MS files (and we expect might be included in future images by future versions of AKF or of the ALMA imaging pipeline). For these latter we suggest a possible definition that might be improved during their development.

* <u>Existing FITS Standard 4.0</u>
    ```
    SIMPLE
    BITPIX
    NAXIS
    NAXISn
    PC##i##j
    Pvi_m
    CTYPEn, CRVALn, CDELTn, CRPIXn, CUNITn
    RADESYS
    SPECSYS
    BSCALE, BZERO,BTYPE
    CHECKSUM
    OBSGEO-X,OBSGEO-Y, OBSGEO-Z
    LONPOLE, LATPOLE
    TELESCOP
    OBJECT
    EQUINOX
    BMAJ,BMIN,BPA
    OBSERVER
    END
    ```

* <u>Data acquisition and reduction keywords</u>
  AKF:
    ```
    KRATAR
    KDECTAR
    KUVRANGE
    KBAND
    KMINPRBL, KMAXPRBL
    KMAXANGS
    KPADLIST
    KNANT
    KMINELT
    KDATEOBS, KDATEEND, KDATEAVG
    KMJDOBS, KMJDEND, KMJDAVG
    KMINEL: source minimum elevation range achieved during observations in 12m, 7m and TP
    data.
    KEXPTIM: exposure time spent on source for each of 12m, 7m and TP array
    KFOV: field of view of each array
    KPROJID: identifier code for the project the image belongs to
    KCASAVER: CASA version(s) used for calibration and imaging processing
    KGOUS: id of the group observing unit set for which the image is one of the products
    KMOUS: id of the member observing unit set for which the image is one of the products
    KSGOAL: id of the science goal for which the image is one of the products
    KSB: id of the scheduling blocks for which the image is one of the products
    KZSOUR: redshift of source
    KQA2FLG: flag given to the member observing unit set during the quality assessment 2
    stage
    ```

* <u>Image description keywords</u>
  AKF
    KBNDRES
    KBNDCTR
    KBNDWID
    KCHNRMS
    KFLUXTOT
    KSPATRES
    KDYNRNG
    KSTOKES
    KDATAMAX, KDATAMIN

## B – Example of suitable FITS header format

In the following, we provide some examples of the format we suggest the FITS keywords should have.

```
EXISTING FITS STANDARD 4.0 KEYWORDS
SIMPLE = T                                  / Standard FITS
BITPIX = -32                                / Floating point (32 bit)
EXTEND =                                                  F
NAXIS = 4                                   / Number of axes in the associated data array.
NAXIS1 = 240                                / NAXIS 1 dimension
NAXIS2 = 240                                / NAXIS 2 dimension
NAXIS3 = 1                                  / NAXIS 3 dimension
NAXIS4 = 1                                  / NAXIS 4 dimension
BSCALE = 1.000000000000E+00                 / PHYSICAL = PIXEL*BSCALE + BZERO
BZERO = 0.000000000000E+00                  / PHYSICAL = PIXEL*BSCALE + BZERO
BTYPE = 'Intensity'                         / Brightness (pixel) unit
BUNIT = 'JY/BEAM '                          / Physical units in which the quantities in array
EQUINOX = 2.000000000000E+03                / Equinox of source coordinates
PC001001 = 1.000000000000E+00               / Transformation matrix terms
PC002001 = 0.000000000000E+00               / Transformation matrix terms
PC003001 = 0.000000000000E+00               / Transformation matrix terms
PC004001 = 0.000000000000E+00               / Transformation matrix terms
PC001002 = 0.000000000000E+00               / Transformation matrix terms
PC002002 = 1.000000000000E+00               / Transformation matrix terms
PC003002 = 0.000000000000E+00               / Transformation matrix terms
PC004002 = 0.000000000000E+00               / Transformation matrix terms
PC001003 = 0.000000000000E+00               / Transformation matrix terms
PC002003 = 0.000000000000E+00               / Transformation matrix terms
PC003003 = 1.000000000000E+00               / Transformation matrix terms
PC004003 = 0.000000000000E+00               / Transformation matrix terms
PC001004 = 0.000000000000E+00               / Transformation matrix terms
PC002004 = 0.000000000000E+00               / Transformation matrix terms
PC003004 = 0.000000000000E+00               / Transformation matrix terms
PC004004 = 1.000000000000E+00               / Transformation matrix terms
PV2_1 = 0.000000000000E+00                  / Parameter value #1 for world coordinate axis #2,
PV2_2 = 0.000000000000E+00                  / Parameter value #2 for world coordinate axis #2,
CTYPE1 = 'RA---SIN'                         / WCS term: type of Axis 1
CRVAL1 = 2.853708750000E+02                 / WCS term: Reference pixel value, axis 1
CDELT1 = -4.444444444444E-05                / WCS term: Increment per pixel, axis1
CRPIX1 = 1.210000000000E+02                 / WCS term: Reference pixel number, axis 1
CUNIT1 = 'deg '                             / WCS term: Unit of axis 1
CTYPE2 = 'DEC—SIN'                          / WCS term: type of Axis 2
CRVAL2 = -3.703011111111E+01                / WCS term: Reference pixel value, axis 2
CDELT2 = 4.444444444444E-05                 / WCS term: Increment per pixel, axis 2
CRPIX2 = 1.210000000000E+02                 / WCS term: Reference pixel number, axis 2
CUNIT2 = 'deg '                             / WCS term: Unit of axis 2
CTYPE3 = 'FREQ '                            / WCS term: type of Axis 3
CRVAL3 = 2.315424966698E+11                 / WCS term: Reference pixel value, axis 3
CDELT3 = 3.870856771975E+09                 / WCS term: Increment per pixel, axis 3
CRPIX3 = 1.000000000000E+00                 / WCS term: Reference pixel number, axis 3
CUNIT3 = 'Hz '                              / WCS term: Unit of axis 3
CTYPE4 = 'STOKES '                          / WCS term: type of Axis 4
CRVAL4 = 1.000000000000E+00                 / WCS term: Reference pixel value, axis 4
CDELT4 = 1.000000000000E+00                 / WCS term: Increment per pixel, axis 4
```

```
CRPIX4 = 1.000000000000E+00               / WCS term: Reference pixel number, axis 4
CUNIT4 = ' '                              / WCS term: Unit of axis 4
RADESYS = 'ICRS '                         / Reference system for equatorial coordinates
RESTFRQ = 2.315424966698E+11              / Rest Frequency (Hz)
SPECSYS = 'LSRK'                          / Spectral reference frame
OBSGEO-X=    2.225142180269E+06           / [m] X  coordinate of observation position wrt
                                             Geocentric reference
OBSGEO-Y=   -5.440307370349E+06           / [m] Y  coordinate of  observation  position  wrt
                                             Geocentric reference
OBSGEO-Z=   -2.481029851874E+06           / [m] Z  coordinate  of  observation  position  wrt
                                             Geocentric reference
LONPOLE =    1.800000000000E+02           / [deg] Long. in native coordinate system of
                                             celestial system's north pole
LATPOLE =   -7.667444444444E-01           / [deg] Lat in native coordinate system of celestial
                                             system's north pole
RA = 2.853708750000E+02                   / [deg] Image centre RA
DEC = -3.703011111111E+01                 / [deg] Image centre Dec
OBJECT  = 'Mrk590'                        / target name
TELESCOP= 'ALMA '                         / Telescope name
OBSERVER= 'koayjy   '                     / Alma id of the PI
BMAJ = 2.228875623809E-04                 / [arcsec] Restoring beam  major axis
BMIN = 1.697528362274E-04                 / [arcsec] Restoring beam  minor axis
BPA = 5.713778686523E+01                  / [deg] Restoring beam position angle
TIMESYS = 'UTC '                          / Time system for time-related kwds in the HDU
DATE = '2012-10-11T09:27:32.760000'       / Date FITS HDU file was written
ORIGIN  = 'JAO'                           / Organization responsible for producing dataset.
END                                       / End of HDU

AKF IMAGE KEYWORDS
KDATAMAX = 1.5                            / [Jy/beam] Maximum value in the FITS image
KDATAMIN = -.05                           / [Jy/beam] Minimum value in the FITS image
KDYNRNG = 5.0                             / Dynamic range in the image
KCHNRMS = 0.0003                          / [Jy/beam] RMS per channel of FITS image
KSPATRES = 0.7                            / [arcsec] Spatial resolution of the FITS image
KBNDCTR = 2.315424966698E+11              / [Hz] Center frequency of data in the FITS array
KBNDWID =1.875E+9                         / [Hz] Bandwidth of the FITS image
KBNDRES =0.488281E+6                      / [Hz] Frequency resolution in the FITS image
KSTOKES = 'I '                            / List of data Stokes parameters

AKF DATA ACQUISITION AND REDUCTION KEYWORDS
KDATEOBS = '2012-06-17T05:56:15.792000'   / Date and time of start of observations
KDATEAVG = '2012-06-17T15:56:15.792000'   / Mid-point of the observations
KMJDOBS = 55927.50000                     / Modified Julian Date of start of the observation
KMJDAVG =55928.54321                      / Modified JD of the mid-point of observation
KMJDEND= 55929.0321                       / Modified JD of the last observation
KDATEEND = '2012-06-18T05:56:15.792000'   / Date and time of last observations
KRATARG = 2.853708750000E+02              / [deg] RA of the source
KDECTARG = -3.703011111111E+01            / [deg] Dec of the source
KBAND='B03'                               /  Observing band of the observation
KMAXANGS = 10.19                          / [arcsec] Maximum ang. scale of the FITS image
KUVRANGE='{''Q3'':160.26489980944029,''Q1'':'67.647891461506305,''MEDIAN'':
106.00663645697135}'                      / [klambda] Median, 1st and 3rd quartile of the UV
                                             length distribution in klambda
KMINPRBL='15.0535428486'                  / [m] Minimum baseline
KMAXPRBL='327.791049213'                  / [m] Maximum baseline
KPADLIST='[''A137'',''A040'',''A068'',''A030'',''A058'',''A070'',''A043'',''A071'',''A013''
, ''A035'', ''A019'', ''A017'']'          / List of ALMA pad names contributing to data
KNANT='{''7M'':[0],''12M'':[31],''TP'':[0]}'
                                          / Number of ALMA 7m, 12m and TP antennas used
KMINEL='{''7M'': None,''12M'':[-22.76],''TP': None}'
                                          / [deg] Minimum elevation of ALMA  12 m  antennas
KFOV = 0.1234                             / [deg^2] Total field of view of the image
KEXPTIM='{''t7M'': None,''t12M'':[324],''tTP': None}
                                          / [s] On-source observing time of ALMA 12m, 7m, and
                                             TP antennas

ADDITIONAL KEYWORDS
KZSOURC = 0.000485787                     / Redshift of source
KPROJID = ''2011.0.00101.S'               / ALMA proposal ID
KCASAVER = 'CASA 3.4.0(release r19988)'   / Version of CASA used to produce FITS image,
                                             i.e. the output of casa-config --version
```

```
KGOUS  = 'uid://A005/X006/X007'         / Group observing unit set ID
KMOUS  = 'uid://A008/X009/X010'         / Member observing unit set status ID
KSOUS  = 'uid://A009/X010/X011'         / Science Goal Observing unit set ID
KSB    = 'exampleSB1 '                  / Names of scheduling blocks contributing to data in
                                          array
KQA2FLG = 'PASS'                        / QA2 flag description
```

## C – Example of scientific use of AKF task

In this Appendix, we show one possible scientific use of the AKF CASA task (Sect. 4). In particular, we report the Python code to produce the light curves shown in Fig.6. The Python code is also available in the IT ARC web site on the AKF page[7].

**Light curves Python code:**

```
from astropy.io import fits
import numpy as np
import matplotlib.pyplot as plt
import ast

kwdlist_lc=['KMJDAVG', 'KFLUXTOT', 'KCHNRMS']
tvar=[]
tvar_fff=[]
fvar=[]
tvar2=[]
tvar2_fff=[]
fvar=[]
z=0.651
images_b3 = ['/b3/1/J0635-7516.fits', '/b3/2/J0635-7516.fits', '/b3/3/J0635-7516.fits']

for i in images_b3:

        AKF(imName=i,kwdlist=kwdlist_lc, include=True)
            hdul = fits.open(i)

        tvar_f= hdul[0].header['KMJDAVG']
        tvar_ff=float(tvar_f)
        tvar_min=min(tvar)
        tvar_fff.append(float((tvar_ff-tvar_min)/(1+z)))

        fvar_f= hdul[0].header['KFLUXTOT']
        fvar_s= ast.literal_eval(fvar_f)
        fvar_p= fvar_s['I']
        fvar_m= log(fvar_p, 10)
        chnrms= hdul[0].header['KCHNRMS']
        chnrms_s= ast.literal_eval(chnrms)
        chnrms_p= chnrms_s['I']
        efvarb3=5*fvar_p/100 + chnrms_p
        lefvarb3= efvarb3/(fvar_p*log(10))
        fvar.append(float(fvar_m))

images_b6 = ['/b6/1/J0635-7516.fits', '/b6/2/J0635-7516.fits', '/b6/3/J0635-7516.fits']

for l in images_b6:

        AKF(imName=l,kwdlist=kwdlist_lc, include=True)
            hdul = fits.open(i)

        tvar2_f= hdul[0].header['KMJDAVG']
        tvar2_ff=float(tvar2_f)
        tvar2_min=min(tvar2)
        tvar2_fff.append(float((tvar2_ff-tvar2_min)/(1+z)))
```

---

7   http://www.alma.inaf.it/index.php/ALMA_FITS_Keywords

```
        fvar2_f= hdul[0].header['KFLUXTOT']
        fvar2_s= ast.literal_eval(fvar2_f)
        fvar2_p= fvar2_s['I']
        fvar2_m= log(fvar2_p, 10)
        chnrms2= hdul[0].header['KCHNRMS']
        chnrms2_s= ast.literal_eval(chnrms2)
        chnrms2_p= chnrms2_s['I']
        efvarb6=5*fvar2_p/100 + chnrms2_p
        lefvarb6= efvarb6/(fvar2_p*log(10))
        fvar2.append(float(fvar2_m))
plt.figure()
plt.title('J0635-7516')
plt.xlabel('t[days]')
plt.ylabel('Log(S[Jy]')
plt.errorbar( tvar_fff,   fvar, yerr=lefvarb3, label="band 3", fmt="bs", linewidth=3)
plt.errorbar( tvar2_fff,  fvar2, yerr=lefvarb6,  label="band 6", fmt="rs", linewidth=3)
plt.ylim(-0.5,0.5)
plt.legend()
plt.show()
```

## Acknowledgements

We would like to thank the referee Peter Teuben for his useful comments. This manuscript and the AKF and KAFE development are part of the activities for the ALMA re-imaging Study approved in the framework of the 2016 ESO Call for Development Stud- ies for ALMA Upgrade (PI: Massardi).